\documentclass[10pt,journal,compsoc]{IEEEtran}
\ifCLASSOPTIONcompsoc
  \usepackage[nocompress]{cite}
\else
  \usepackage{cite}
\fi

\ifCLASSINFOpdf

\else

\fi

\usepackage{amsmath,amssymb,amsfonts}
\usepackage[inline]{enumitem}
\usepackage{graphicx}
\usepackage{caption}
\usepackage{subcaption}
\usepackage{amsmath}
\usepackage{algorithm}
\usepackage{algpseudocode}
\usepackage{multicol}
\usepackage{url}
\usepackage{hyperref}
\usepackage{xcolor}

\begin{document}

\title{A Deep Recurrent-Reinforcement Learning Method for Intelligent AutoScaling of Serverless Functions}
\author{Siddharth~Agarwal,
        Maria~A.~Rodriguez,
        and~Rajkumar~Buyya%        
\IEEEcompsocitemizethanks{\IEEEcompsocthanksitem The authors are with the Cloud Computing and Distributed Systems (CLOUDS) Laboratory, School of Computing and Information Systems, The University of Melbourne, VIC 3010, Australia.\protect\\
E-mail: \href{mailto:siddhartha@student.unimelb.edu.au}{siddhartha@student.unimelb.edu.au},\href{mailto:maria.read@unimelb.edu.au}{maria.read@unimelb.edu.au} ,\href{mailto:rbuyya@unimelb.edu.au}{rbuyya@unimelb.edu.au}}% 
}

\IEEEtitleabstractindextext{%
\begin{abstract}
Function-as-a-Service (FaaS) introduces a lightweight, function-based cloud execution model that finds its relevance in a range of applications like IoT-edge data processing and anomaly detection. While cloud service providers (CSPs) offer a near-infinite function elasticity, these applications often experience fluctuating workloads and stricter performance constraints. A typical CSP strategy is to empirically determine and adjust desired function instances or resources, known as \textit{autoscaling}, based on monitoring-based thresholds such as CPU or memory, to cope with demand and performance. However, threshold configuration either requires expert knowledge, historical data or a complete view of the environment, making autoscaling a performance bottleneck that lacks an adaptable solution. Reinforcement learning (RL) algorithms are proven to be beneficial in analysing complex cloud environments and result in an adaptable policy that maximizes the expected objectives. Most realistic cloud environments usually involve operational interference and have limited visibility, making them partially observable. A general solution to tackle observability in highly dynamic settings is to integrate Recurrent units with model-free RL algorithms and model a decision process as a Partially Observable Markov Decision Process (POMDP). Therefore, in this paper, we investigate model-free Recurrent RL agents for function autoscaling and compare them against the model-free Proximal Policy Optimisation (PPO) algorithm. We explore the integration of a \textit{Long-Short Term Memory} (LSTM) network with the state-of-the-art PPO algorithm to find that under our experimental and evaluation settings, recurrent policies were able to capture the environment parameters and show promising results for function autoscaling. We further compare a PPO-based autoscaling agent with commercially used threshold-based function autoscaling and posit that a LSTM-based autoscaling agent is able to improve throughput by 18\%, function execution by 13\% and account for 8.4\% more function instances.
	\end{abstract}

\begin{IEEEkeywords}
Serverless Computing, Function-as-a-Service, AutoScaling, Reinforcement learning, Constraint-awareness 
\end{IEEEkeywords}}

\maketitle
\IEEEdisplaynontitleabstractindextext
\IEEEpeerreviewmaketitle
	
	\ifCLASSOPTIONcompsoc
\IEEEraisesectionheading{\section{Introduction}\label{sec:introduction}}
\else
	\section{Introduction}
 \label{sec:introduction}
\fi
\IEEEPARstart{T}{h}e growing popularity of event-driven application architectures fuel the increased adoption of serverless computing platforms. Serverless computing introduces a cloud-native execution model that offloads server governance tasks to the cloud service provider (CSP) and aims to reduce operational costs. Serverless features a variety of attributes like a microservices-inspired architecture, high elasticity, usage-based resource billing, and zero idle costs. Function-as-a-Service (FaaS) is a function-based abstraction of serverless computing that decouples an application into functions, small pieces of business logic, that execute on a lightweight virtual machine (VM) or container. These functions generally serve a single purpose, run for a very short duration, and do not maintain a state to enable faster scaling \cite{qu2018auto}. Functions can be associated with multiple event sources such as HTTP events, database or storage events, and IoT notifications that execute function handlers or business logic and respond to incoming workloads.  

Serverless, often used interchangeably with FaaS, has attracted a wide range of application domains such as IoT services, REST APIs, stream processing, and prediction services. These applications may have strict availability and QoS requirements, i.e., throughput and response time while having fluctuating resource requirements that uniquely affect function performance. To address performance constraints and handle complex workloads, FaaS platforms heuristically spin up a new function instance, i.e., function autoscaling, for each incoming request and shut down the instance after service \cite{jonas2019cloud} to free up resources. However, FaaS offerings such as AWS Lambda, Azure Functions, Google Cloud Functions, OpenFaaS \cite{openfaas} and Kubeless \cite{kubeless} may choose to re-use a function instance or keep the instance running for a limited time to serve subsequent requests \cite{vahidinia2020cold}. A recent study  \cite{manner2022resource} asserts that appropriate resource allocation, i.e., CPU and memory, is needed to guarantee QoS fulfillment and improve business value in serverless computing. \textit{Autoscaling} is the process of adding or removing function(s) from a platform, as per the demand, and has a direct correlation with platform performance. CSPs usually employ general-purpose rule-based or threshold-based horizontal scaling mechanisms or utilize a pool of minimum running function(s) \cite{lamdascaling}\cite{gcfscaling} to handle function start-up delays while serving workload.    

Autoscaling provides an opportunity for CSPs to optimally utilize their resources \cite{gari2021reinforcement} and share unused resources in a multi-tenant environment. However, configuring thresholds involves manual tuning, expert domain knowledge, and application context that reduces development flexibility and increases management overhead. Since cloud workloads are highly dynamic and complex, threshold-based autoscaling solutions lead to challenges like function cold starts and hysteresis \cite{jawaddi2023autoscaling}, failing to offer performance guarantees. A \emph{cold start} is a non-negligible function instantiation delay that is introduced before processing the request, while hysteresis highlights the temporal dependency of environment states on the past. Therefore, providing an adaptive, flexible, and online function autoscaling solution is an opportunity to ensure efficient resource management with performance trade-offs in serverless computing. Furthermore, autoscaling approaches employed by existing FaaS frameworks are excessively dependent on monitoring solutions. Although researchers in \cite{straesser2022not} identify metric collection for thresholds as a bottleneck for autoscaling due to significant collection delay or unreliability, a self-corrective model is demanded to account for underlying variations.  

Autoscaling has been actively investigated in the cloud computing domain \cite{qu2018auto}\cite{jawaddi2023autoscaling}\cite{nguyen2020horizontal}\cite{xu2019adaptive}\cite{xu2022coscal}, particularly for VMs, and has periodically highlighted the need for appropriate resource scaling to minimize operational costs and improve performance. Resource scaling is an NP-hard problem \cite{gari2021reinforcement}\cite{xu2022coscal} and necessitates the realization of complex environmental factors while balancing the system performance between QoS and SLAs. In the past, Reinforcement Learning (RL) algorithms have been applied in the context of VM autoscaling \cite{gari2021reinforcement}\cite{jawaddi2023autoscaling}\cite{straesser2022not}\cite{benedetti2022reinforcement} and have demonstrated adaptable performance over traditional methods in capturing the workload uncertainty and environment complexity. But the application of RL for function autoscaling is yet underexplored \cite{benedetti2022reinforcement}. RL-based solutions are known to interact with an environment, perform an action, learn periodically through feedback, and account for the dynamics of the cloud environment. 

In this work, we investigate the application of Recurrent Neural Networks (RNN), specifically Long-Short Term Memory (LSTM) in a model-free Partially Observable Markov Decision Process (POMDP) setting for function autoscaling. Earlier works \cite{jawaddi2023autoscaling}\cite{zafeiropoulos2022reinforcement}\cite{agarwal2021reinforcement}\cite{schuler2021ai} employing RL-based autoscaling generally model decision making as Markov Decision Process (MDP) and fall short to discuss partial observability in realistic environments \cite{ni2021recurrent}\cite{hausknecht2015deep}. Furthermore, various existing studies discussed in \cite{jawaddi2023autoscaling}\cite{zafeiropoulos2022reinforcement} experiment with RL-based solutions in a simulated FaaS environment, with the research in \cite{straesser2022not} criticizing this methodology. Simulated FaaS frameworks generally sample factors such as cold start and execution time from profiled data and are insufficient to capture the variability in real environments. Therefore, we examine the integration of LSTM with Proximal Policy Optimization (PPO), a state-of-the-art RL algorithm, to analyze partial observability and sequential dependence of autoscaling actions and find a balance between conflicting CSP and user objectives. We perform experiments with \textit{matrix multiplication} function and compare LSTM-PPO against Deep Recurrent Q-Network (DRQN) and PPO (clipped objective) to infer that in our experimental settings, recurrent policies capture the environment uncertainty better and showcase promising performance in comparison to PPO and commercially adopted threshold-based approaches. We make use of OpenAI Stable Baseline's\cite{stable-baselines3} standard implementation of the LSTM-PPO and PPO algorithms, and implement our compatible OpenFaaS serverless environment following Gymnasium \cite{towers_gymnasium_2023} guidelines.

In summary, the key contributions of our work are:
\begin{enumerate}
\item We analyze the characteristics of FaaS environments to identify and model autoscaling decisions as a POMDP. We further hypothesise that scaling decisions have a sequential dependence on interaction history. We propose a POMDP model that captures function metrics such as CPU and memory utilization, function replicas, average execution time and throughput ratio, as partial observations and formulate the scaling problem. 
\item We investigate how function autoscaling works, highlight the differences between contrasting approaches and investigate a Deep Recurrent RL (LSTM-PPO) autoscaling solution to capture the temporal dependency of scaling actions and workload complexity. We deploy the proposed agent to the OpenFaaS framework and utilise open-source function invocation traces \cite{shahrad2020serverless} from a production environment to perform experiments with a matrix multiplication function.
\item We implement a Gymnasium \cite{towers_gymnasium_2023} compatible OpenFaaS serverless environment to be integrated directly with the proposed RL agent.
\item We perform our experiments on Melbourne Research Cloud (MRC) and evaluate the proposed LSTM-PPO approach against the state-of-the-art PPO algorithm, commercially offered threshold-based horizontal scaling, OpenFaaS' request-per-second scaling policy, and a Deep Recurrent Q-Network i.e., DRQN, to demonstrate LSTM-PPO's ability to capture environment uncertainty for efficient scaling of serverless functions.
\end{enumerate}

The rest of the paper is organised as follows. Section \ref{section2} highlights related research studies. In Section \ref{section3}, we present the system architecture and formulate the problem statement. Section \ref{section4} outlines the proposed agent’s workflow and describes the implementation hypothesis and assumptions. In Section \ref{section5}, we evaluate our technique with the baseline approaches and highlight training results and discuss performance. Section \ref{section6} concludes the paper and highlights future research directions.
% 

% \usepackage{graphicx}
% \setlength\extrarowheight{4pt}
 \begin{table*}[t!]
\caption{A Summary of Related Works and Their Comparison with Our Proposed Method. \\
    H: Horizontal Scaling, V: Vertical Scaling}
\label{tab:relatedwork}
\resizebox{\linewidth}{!}{
\begin{tabular}{c| c|c|c|c|c}
\hline
\textbf{Work} &
  \textbf{Type} &
  \textbf{Scaling} &
  \textbf{Technique} &
  \textbf{Objective} &
  \textbf{Environment} \\ \hline
\cite{lamdascaling} & FaaS & H & Threshold-Based        & CPU Utilisation            & AWS Lambda              \\
\cite{gcfscaling} & FaaS & H & Threshold-Based        & CPU Utilisation            & Google Cloud Functions              \\
\cite{xu2022coscal} & Microservices & H,V,Brownout & GRU + Q-Learning        & QoS            & Testbed              \\
\cite{benedetti2022reinforcement} & FaaS          & H            & Q-Learning              & QoS            & OpenFaaS             \\
\cite{zafeiropoulos2022reinforcement} & FaaS          & H            & Q-Learning, DQN, DynaQ+ & QoS + Budget   & Simulation, Kubeless \\
\cite{agarwal2021reinforcement} & FaaS          & H            & Q-Learning              & QoS            & Kubeless             \\
\cite{schuler2021ai} & FaaS          & H            & Q-Learning              & QoS            & Knative              \\
\cite{phung2022prediction} & FaaS          & H            & Bi-LSTM                 & Resource       & Knative              \\
\cite{zhang2022adaptive} & FaaS          & H, V         & Q-Learning              & QoS + Resource & Testbed              \\
\cite{li2022kneescale} & FaaS          & H            & Kneedle Algorithm       & QoS + Budget   & OpenFaaS             \\
Our Method & FaaS          & H            & LSTM - PPO              & QoS + Resource & OpenFaaS\\ \hline
\end{tabular}}
	\vspace*{-3.5mm}
\end{table*}

\section{Related work} \label{sec_related}
 In this section, we summarise (see Table \ref{tab:relatedwork}) existing work on serverless computing, autoscaling in FaaS, and the application of RL in FaaS. We compare existing work based on their key features and provide a detailed background on the Deep Recurrent RL (RPPO) algorithm used in designing our autoscaling policy.

\label{section2}
\subsection{Serverless Computing and Function-as-a-Service}

Serverless computing puts forward a cloud service model wherein the server management or resource management responsibility lies with the CSP. In \cite{jonas2019cloud}, the authors discuss the potential of this new, less complex computing model introduced by Amazon in 2014. The study briefly explains a function-based, serverless commercial offering of AWS Lambda, i.e., the Function-as-a-Service platform. It highlights three primary differences between traditional cloud computing and serverless computing – decoupled computation and storage, code execution without resource management, and paying in proportion to the resources used. The research posits that the serverless or FaaS model promotes business growth, making the use of the cloud easier.

Baldini \emph{et al.} \cite{baldini2017serverless} introduce the emerging paradigm of FaaS as an application development architecture that allows the execution of a piece of code in the cloud without control over underlying resources. The research identifies containers and the emergence of microservices architecture as the promoter of the FaaS model in serverless. The study uses FaaS and serverless interchangeably and defines it as a ‘stripped down’ programming model that executes stateless functions as its deployment unit.

Since the inception of serverless computing, there have been many commercial and open-source offerings such as AWS Lambda, Microsoft Azure Functions, Google Cloud Functions, Fission, and OpenWhisk. These platforms represent FaaS as an emerging technology, but Hellerstein \emph{et al.} \cite{hellerstein2018serverless} put together gaps that furnish serverless as a bad fit for cloud innovations. The authors criticize the current developments of cloud computing and state that the potential of cloud resources is yet to be harnessed. On the contrary, the researchers in \cite{shafiei2019serverless} argue that the serverless offerings are economical and affordable as they remove the responsibility of resource management and complexity of deployments from consumers. They discuss the opportunities offered by multiple FaaS offerings and give an overview of other existing challenges, and indicate potential approaches for future work. 

In an article by Microsoft \cite{rosenbaum}, Rosenbaum estimates that there will be nearly 500 million new applications in the subsequent five years, and it would be difficult for the current development models to support such large expansions. FaaS is designed to increase development agility, reduce the cost of ownership, and decrease overheads related to servers and other cloud resources. The term 'serverless' has been in the industry since the introduction of Backend-as-a-Service (BaaS). Despite the serverless benefits, FaaS experiences a few challenges, categorized as system-level, and programming and DevOps challenges \cite{jonas2019cloud}\cite{baldini2017serverless}\cite{rosenbaum}. The former identifies the cost of services, security, resource limits, and cold start while scaling, and the latter focuses on tools and IDEs, deployment, statelessness, and code granularity in the serverless model.

\subsection{AutoScaling in Function-as-a-Service}

Resource elasticity, analogously used with autoscaling, is a vital proposition of cloud computing that enables large-scale execution of a variety of applications. A recent survey \cite{gari2021reinforcement} discusses the relevance of cloud resource elasticity for the Infrastructure-as-a-service (IaaS) model to express that autoscaling and pay-as-you-go billing enables infrastructure adjustments based on workload variation while complying with SLAs. On this basis, the study identifies that autoscaling addresses a set of associated challenges, namely, scaling and scheduling which are generally NP-hard problems. Additionally, the research explores the possibility of RL algorithms for autoscaling to approach the complexity and variability of cloud environments and workloads. It is emphasized that utilization of such RL algorithms for scaling purposes can help the service providers to come up with a more transparent, dynamic, and adaptable policy.

Straesser et al. \cite{straesser2022not} conduct experiments related to cloud autoscaling and assert autoscaling to be an important aspect of computing for its effects on operational costs and QoS. The authors define scaling as a task of dynamically provisioning resources under a varying load and necessitates the automation of processes for highly complex cloud workloads. They discuss that commercial solutions usually operate with user-defined rules and threshold heuristics, and state that an optimal autoscaler is expected to minimize operational cost and SLA violations. 

In addition to workload variability, QoS sensitivity is also identified as an enabler for increased operational costs and resource wastage. A microservices-focused autoscaling scheme is introduced in \cite{xu2022coscal} where a trade-off between horizontal, vertical, and a self-adaptable brownout technique is determined based on the infrastructure and workload conditions. The researchers exploit Gated-Recurrent Units (GRUs) for workload prediction and utilize Q-learning for making trade-off updates and scaling decisions. The study asserts that workload prediction is an important factor for autoscaling and acknowledges resource allocation to be an NP-hard problem with multi-dimensional objectives of QoS and SLAs. 

In the context of FaaS autoscaling, work in \cite{schuler2021ai} experiments with the concurrency-level setting of Knative, a Kubernetes-based serverless framework, and identify that function concurrency settings have varying effects on latency and throughput of function. Therefore, they utilize the Q-learning algorithm to configure functions with optimal concurrency levels to further improve performance. Another work \cite{agarwal2021reinforcement} presents preliminary results of applying Q-learning to FaaS for predicting the optimal number of function instances to reduce the cold start problem. They utilize the function resource metrics and performance metrics and apply them to discrete state and action spaces for adding or removing the function replicas, with threshold-based rewards, to eventually improve function throughput.

Similarly, studies like \cite{zafeiropoulos2022reinforcement}\cite{phung2022prediction}\cite{zhang2022adaptive} emphasize addressing the dynamicity, agility, and performance guarantees of FaaS by employing RL-based autoscaling solutions. The work in \cite{zafeiropoulos2022reinforcement} follows a monitoring-based scaling pattern and explores algorithms like Q-learning, DynaQ+, and Deep QL, partially in simulation and practical settings, to reasonably utilize resources and balance between budget and QoS. They aid the agent's training process by sampling simulation data based on probability distribution and running parallel agents to speed up the learning process. The work in \cite{phung2022prediction}, discusses the concurrency level in the Knative framework and asserts that identifying appropriate thresholds is challenging, requires expert knowledge, and has varying effects on performance. Therefore, to efficiently use the function resources and improve performance, authors profile different concurrency levels for best performance and propose an adaptive, Bi-Long Short Term Memory (Bi-LSTM) model for workload prediction and determine the number of function replicas using identified concurrency levels. Another study \cite{zhang2022adaptive} focuses on function response time and states that threshold-based scaling cannot devise a balance between resource efficiency and QoS. Therefore, the authors explore Q-learning to propose adaptive horizontal and vertical scaling techniques by profiling different resource allocation schemes and their corresponding performance. Their proposed state space considers resource requests and limits, along with the availability of GPU components, to model rewards as the divergence from agreed SLO levels. Taking a different approach, the researchers in \cite{benedetti2022reinforcement} utilize Q-learning in the context of Kubernetes-based serverless frameworks and propose a resource-based scaling mechanism to adjust function CPU utilization threshold to reduce response time SLA violations. Taking a different approach, \cite{li2022kneescale} proposes an online application profiling technique that identifies a knee point and adjusts resources until the point those changes reflect in performance gain using the Kneedle algorithm in conjunction with binary search. Further, a survey \cite{jawaddi2023autoscaling} summarises autoscaling techniques for serverless computing under different categories like rule-based, AI-based, analytical model, control theory-based, application profiling, and hybrid technique and envisions new directions like energy-driven and anomaly-aware serverless autoscaling.

These proposals are complementary yet contrasting to each other either in optimization objectives, profiled metrics, or scaling policy. Some fail to address the performance dependency on complex workloads, while few rely on pre-configured thresholds \cite{lamdascaling}\cite{gcfscaling} that require expert knowledge and application insights. Few studies focusing on workload prediction assume a fully observable environment and miss out on the temporal dependency of environment states where scaling decisions have been taken. Contradictory to these proposals, we examine a Deep Recurrent RL-based autoscaling solution, particularly LSTM-PPO, to hypothesize that FaaS environments are highly dynamic, partially observable with complex workloads, and that scaling decisions are influenced by environment uncertainty. We model function autoscaling as a partially observable Markov decision process (POMDP) and utilize monitoring metrics like average CPU and memory utilization, function resource requests, average execution time, and throughput ratio to discover an optimal scaling policy. Our proposed RL-based autoscaling agent interacts with the FaaS environment, waits for a sampling period \cite{straesser2022not} to receive delayed rewards, and feeds the observed environment state to the recurrent actor-critic model. Although a few studies \cite{xu2022coscal}\cite{phung2022prediction} have utilized recurrent networks like LSTM or GRU for workload prediction in serverless context but do not address the temporal relationship between scaling actions and their effect on environment state. Further, we take inspiration from \cite{hausknecht2015deep}\cite{gold2003fx}\cite{hu2022effective} where recurrent models have been utilized to analyze the inter-dependence of environment states and retain useful information to learn optimal policies.

\begin{figure*}[t!]
	\centering 
	\includegraphics[width=0.85\linewidth]{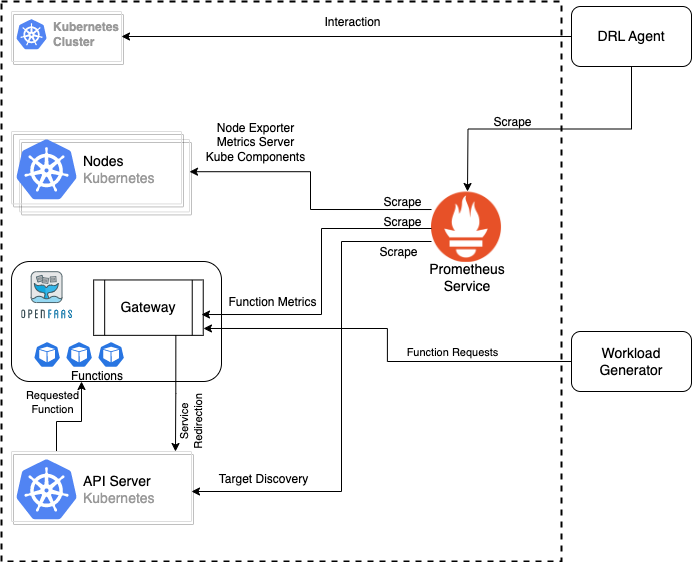}
	\vspace*{-2.0mm} 
	\caption{System Architecture}
	\label{fig:systemArchitecture}
	\vspace*{-4mm}
\end{figure*}

\section{System Architecture and Problem Formulation}
\label{section3}

\subsection{System Architecture}
The main components of our autoscaling solution are the Prometheus monitoring service and the DRL agent, which are shown in Fig. \ref{fig:systemArchitecture}. For the serverless environment, we deploy OpenFaaS \cite{openfaas}, a Kubernetes-based FaaS framework, over a multi-node MicroK8s \cite{microk8s} cluster, a production Kubernetes distribution. OpenFaaS includes a Gateway deployment to expose function performance metrics and Prometheus is configured to periodically scrape function metrics such as execution time, replica count, and throughput ratio. OpenFaaS also packs an \textit{alertmanager} that periodically watches for pre-configured request-per-second scaling threshold to provide horizontal scaling capabilities. The monitoring service further scrapes resource metrics from the Kubernetes API Server, Kubelet, and Node exporters that are utilized by our DRL agent for observation collection at every sampling window. The DRL agent utilizes the standard Stable Baseline3 (SB3) \cite{stable-baselines3} implementation of LSTM-PPO \footnote{\label{gitrepo}\url{https://github.com/Cloudslab/DRe-SCale}} and models the FaaS environment following Gymnasium \cite{towers_gymnasium_2023} guidelines, for the POMDP model to be directly used by SB3 algorithms. Additionally, we implement our own version of DRQN \footnotemark[1] using PyTorch \cite{drqn}\cite{dqnagents} for evaluation. We also deploy an HTTP-request generator tool to simulate online user behavior to train and evaluate our DRL autoscaling agent.

\subsection{Problem Formulation}
\label{sec:problem_formulation}
Existing FaaS platforms generally exercise threshold-based scaling when a monitored metric exceeds the configured maximum or minimum. Autoscaling of resources is considered a classic automatic control problem and commonly abstracted as a MAPE (map-analyse-plan-execute) control loop \cite{qu2018auto}. At every sampling interval, the monitoring control loop collects the relevant metrics and may decide to scale based on the analyzed observation. Autoscaling is a sequential process with non-deterministic results in a partially observable environment that is conditioned on historical interactions, therefore, we design FaaS autoscaling as a \textit{model-free POMDP}. POMDPs are a mathematical model and an extension of Markov Decision Processes (MDP) that account for uncertainty while maximising a given objective.

\subsubsection{Model-Free POMDP}
\label{sub:pomdp}

In a real-world scenario, it is hard to perceive the complete state of the surrounding environment and a MDP rarely holds true \cite{ni2021recurrent}. Instead, a POMDP better encapsulates environmental characteristics from incomplete or partial information about said environment. Formally, a POMDP model is defined as a 6-tuple ($S,A,O,T,Z,R$) where: {$S$} denotes the set of all possible environment states, {$A$} denotes the set of all actions, {$O$} denotes the set of all observations that an agent can perceive, {$T$} and {$Z$} represent the transition probability function and observation probability function, respectively, and {$R$} denotes the reward function. Conceptually, the agent observes itself in some environment state {$s_t$}, hidden due to partial observability at each sampling interval {$t$} and maintains a belief {$b_t$}, an estimate of its current state, to select an action {$a_t$} and transition to a new state {$\hat{s}_t$}. The agent perceives the state information through observation {$o_t$} and utilises the transition and observation probability function to update the state estimates. After transitioning to a new state {$\hat{s}_t$}, the agent receives reward {$r_t$} that helps in maximising the objective. 

Since probability functions are difficult to model in complex FaaS environments and states cannot be perfectly represented to capture the estimates of belief or hidden states \cite{hausknecht2015deep}, we define the autoscaling problem as model-free POMDP. Model-free POMDP attempts to maximise the cumulative reward without explicitly modelling the transition or observation probabilities. Further, it needs function approximation techniques like neural networks, specifically recurrent neural networks (RNN), to capture the uncertainty and temporal dependency. Therefore, we define the POMDP observations as a tuple of \textit{$(O, A, R)$} and utilise recurrency to model and infer transition probabilities, observation probabilities and hidden states to fulfil the conflicting objectives of resource utilisation, operational cost and QoS objectives.

\begin{table}[!t]
\caption{Notations}
\label{tab_notation} 
\centering 
\footnotesize
\begin{tabular}{p{1.7cm}p{6.0 cm}} 
\hline
 Symbol & Definition\\
\hline
  $f_l$ & Function for training \{matmul\}. \\
  $S$ & State space for POMDP agent. \\
  $A$ & Action space for POMDP agent. \\
  $O$ & Observation space for POMDP agent. \\
  $T, Z$ & Transition and Observation probability functions.\\
  $R$ & Reward function for POMDP agent.\\
  $N$ & Maximum number of function replicas possible (function quota). \\
  $n_{min}$ & Minimum number of function replicas.  \\
  $Q$ & Maximum requests possible in a sampling window. \\
  $t$ & sampling window. \\
  $n_t$ & All available functions during $t$. \\
  $\tau_t$ & Average execution time of $n_t$ functions. \\
  $c$ & Average CPU utilisation of $n_t$ functions.\\
  $c_{max}$ & Maximum CPU utilisation of a function. \\
  $m_t$ & Average memory utilisation of $n_t$ functions. \\
  $m_{max}$ & Maximum memory utilisation of a function. \\
  $\phi_t$ & Throughput of function.\\
  $q_t$ & Requests during $t$. \\
  $k$ & Scaling limits. \\
  $s_t$ & Environment state at $t$. \\
  $b_t$ & Belief state for POMDP agent. \\
  $o_t$ & Environment observation tuple $(\tau_t,\phi_t,q_t,n_t,c_t,m_t) \in$ $O$ at $t$. \\ 
  $a_t$ & Agent action $\in \{-k,\dots,+k\}$ at $t$. \\
  $r_t$ & Reward for action $a_t \in$ $R$ at $t$. \\
  $r_{min}$ & Negative immediate reward (-100).\\
  $\alpha$, $\beta$, $\gamma$ &  Objective weight parameters. \\
 \hline
\end{tabular} 
\vspace{-5.5mm}
\end{table}

\subsubsection{Deep Recurrent-Reinforcement Learning}
\label{subsub:DRL}

A possible solution to learning effective policies in a model-free POMDP is the application of model-free RL algorithms. Here, the agent directly interacts with the environment and does not explicitly model the transition or observation probabilities. Vanilla RL algorithms like Q-learning and DQN have no mechanism to determine underlying state \cite{hausknecht2015deep} and speculates that fed observation is a complete representation of the environment. To capture sequential or temporal dependencies, often recurrent units are integrated with vanilla RL approaches, known as Recurrent Reinforcement Learning (RRL) \cite{li2015recurrent}. Prior studies \cite{hausknecht2015deep}, \cite{gold2003fx} \cite{hu2022effective},\cite{bakker2001reinforcement} \cite{cong2021double}, have introduced and applied RRL approaches to a variety of application domains such as  T-maze task, financial trading, network resource allocation and Atari games, to address sequential nature and partial observability of environment, i.e., a non-Markovian or POMDP setting. In RRL, an agent follows the basic principle of performing an action in the environment, establishing its state and receiving feedback to improve the policy, but, additionally employs RNN units/cells to model uncertainty. Theoretically, POMDP has an underlying dynamics of MDP with an additional constraint of state uncertainty or observability that makes the process non-Markovian. Therefore, we define the core RL components as observation $O$, action $A$, reward $R$ (guiding signals) and FaaS environment.

We model the observation space as $o_t= (\tau_t,\phi_t,q_t,n_t,c_t,m_t)$ $\in O$ where {$\tau_t$} is average execution time of {$n_t$} available function replicas with {$c_t$} average CPU and {$m_t$} average memory utilisation, while successfully serving {$\phi_t$} proportion of {$q_t$} requests in the sampling window {$t$}. The agent adjusts the number of function instances in the upcoming sampling window {$t+1$} using suitable actions in an attempt to maximise the reward. Therefore, we define scaling action {$a_t$} as the number of function instances, {$k$}, to add or remove and represent it as {$a_t$ $\in A = \{-k, \dots +k\}$} such that {${n_{min}} \leq ({n}_{t-1} + a_t) \leq N$}, where {$N$} is function quota. This estimate helps the agent to control the degree of exploration by maintaining replication within quota {$N$}. 

The objective of the DRL agent is to learn an optimal scaling policy, and therefore, we structure the rewards {$r_t$} $\in R$ over monitored metrics - {$c_t$} average CPU utilisation, {$m_t$} average memory utilisation, {$\phi_t$} successful proportion of total requests and number of available function replicas {$n_t$}. Our proposed agent does not work towards achieving a specific threshold. Instead, it learns to maximise the returns, i.e., improve resource utilisation, throughput and economically scaling function replicas. After performing an action {$a_t$}, the agent receives a delayed reward {$r_t$} at every sampling window {$t$} and updates its network parameters. 

RL application for model-free POMDP does not explicitly estimate the probabilities, instead, RNNs are incorporated to analyse environment uncertainties and model time-varying patterns \cite{li2015recurrent}\cite{cong2021double}. The structure of RNNs is made-up of highly-dimensional hidden states that act as network memory and enables it to remember complex sequential data. These networks map an input sequence to output and consist of three units - input, recurrent and output unit, serving towards memory goal.

\begin{figure*}[t!]
	\centering 
	\includegraphics[width=0.8\linewidth]{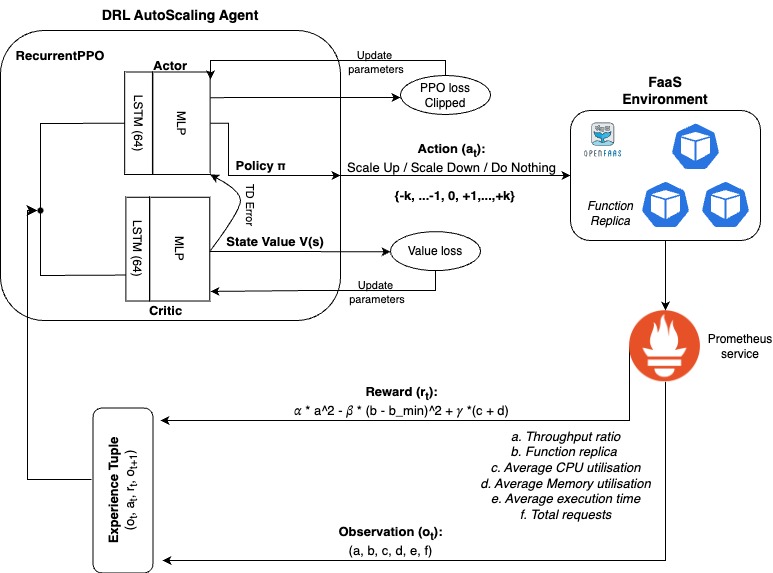}
	\vspace*{-2.0mm} 
	\caption{DRL agent structure for Autoscaling}
	\label{fig:agentFlow}
	\vspace*{-4.5mm}
\end{figure*}

\section{LSTM-PPO based AutoScaling Approach} 
\label{section4}

As discussed in section \ref{sec:problem_formulation}, we introduce recurrency to handle system dynamics, complex workloads, and hidden correlation of components based on POMDP model in autoscaling tasks. We select Proximal Policy Optimisation (PPO), a popular state-of-the-art on-policy RL algorithm for autoscaling agents. While model-free off-policy algorithms such as Deep Q-Network (DQN), Deep Deterministic Policy Gradient (DDPG) have been studied with recurrent units \cite{hausknecht2015deep} \cite{hu2022effective}, we explore a model-free on-policy PPO in our setting due to its ease of implementation, greater stability during learning, better performance across different environments \cite{kozlica2023deep} and support for discrete actions while providing better convergence \cite{stable-baselines3}. Although on-policy methods are known to be sample inefficient and computationally expensive, our agent continuously collects samples for timely policy updates. Also, off-policy algorithms tend to be harder to tune than on-policy because of significant bias from old data and Schulman at el. \cite{DBLP:journals/corr/SchulmanWDRK17} suggests that PPO is less sensitive to hyperparameters than other algorithms. PPO has found its application in domains like robotics, finance and autonomous vehicles, and takes advantage of the Actor-Critic method to learn optimal policy estimations. However, for partial observability or temporal dependence, general RL algorithms struggle to capture underlying correlations and patterns effectively. Therefore, we utilise RNN units, specifically LSTM, to address partial observability in the FaaS environment and improve the agent's decision-making capabilities. This integration is expected to enhance PPO's ability to capture historical data and make informed decisions while improving its policy via new and previous experiences.

The core component of the proposed autoscaling solution is the integration of recurrent units with a fully-connected multi-layer perceptron (MLP) that takes into environment observation and maintains a hidden internal state to retain relevant information. The LSTM layer is incorporated into both actor and critic networks to retain information i.e., the output of the LSTM layer is fed into fully-connected MLP layers, where the actor (policy network) is responsible for learning an action selection policy and the critic network serves as a guiding measure to improve actor's decision. The network parameters are updated as per PPO clipped surrogate objective function \cite{schulman2017proximal} (Eq. \ref{eq:clippedObj}) which helps the agent balance its degree of exploration and knowledge exploitation. It further improves network sample efficiency and conserves large policy updates.

\begin{equation} 
    \label{eq:clippedObj}
    \begin{aligned}
L^{CLIP} (\theta) = \mathbb{\hat{E}_t} \left[\min\left({r_t}(\theta)\hat{A}_t,\text{clip}\left({\hat{r}_t}(\theta), 1 - \epsilon, 1 + \epsilon\right)\hat{A}_t\right)\right]
\end{aligned}
\end{equation}
\vspace{-4mm}
\begin{equation} \label{eq:ratio}
\hat{r}_t(\theta) = \frac{\pi_\theta(a_t|o_t)}{\pi_{\theta_{\text{old}}}(a_t|_t)}
\end{equation}

   \begin{equation}\label{eq:reward}
        r_{t} = 
        \begin{cases}
        \alpha.{\phi_t^{2}} - \beta.{({n_t} - {n_{min}})^{2}} + \gamma.({{c_t} + {m_t}}) & ;\text{$1 \leq a_t + n_{t-1} < N$}\\
        {r_{min}} & ;\text{otherwise}\\
        \end{cases}
    \end{equation}

The proposed autoscaling technique has two phases: an agent training phase and a testing phase. Fig. \ref{fig:agentFlow} demonstrates the agent training workflow. The environment setup process precedes the agent training, where the agent interacts with the environment and obtains information. After initial setup, the agent is trained for multiple episodes of sampling windows, where it assesses the function demand {$q_t$} over individual sampling window {$t$} and ascertains appropriate scaling action. During a sampling window {$t$}, the agent collects the environment observation {$o_t$} and samples an action {$a_t$} according to LSTM-PPO policy. If the agent performs an invalid action, it is awarded an immediate negative reward {$r_{min}$}, else the agent obtains a delayed reward {$r_t$} (Eq. \ref{eq:reward}), for sampling window {$t$}, calculated using the relevant monitored metrics (\ref{sec:problem_formulation}). This reward helps the agent in action quality assessment, transition to a new state and has significant effects on the function's performance. These rewards are essential for improving the agent's decision-making capability. The critic network estimates the agent state and helps update the network parameters. The agent continues to analyse the demand over multiple sampling windows, repeating the interaction process and accumulating the relevant information in recurrent cells for learning. Once the agent is trained for sufficient episodes and rewards appear to converge, we evaluate the agent in the testing phase.

In the testing phase, the agent is evaluated for its learnt policies. It collects current environment observation, samples the action through actor policy and scales the functions accordingly. We hypothesised the relationship between QoS and resource utilisation and deduce that appropriately scaling the functions improve throughput, resource utilisation and reduce operational costs (number of function replicas used).

\section{Performance Evaluation}
\label{section5}
In this section, we provide the experimental setup and parameters, and perform an analysis of our agent compared to other complementary solutions. 

\subsection{System Setup}
We set up our experimental multi-node cluster, as discussed in Section \ref{section3}, using NeCTAR (Australian National Research Cloud Infrastructure) services on the Melbourne Research Cloud. It includes a combination of 2 nodes with 12/48, 1 node with 16/64, 1 node with 8/32 and 1 node with 4/16 vCPU/GB-RAM configurations. We deploy OpenFaaS along with Prometheus service on MicroK8s (v1.27.2), however, we used Gateway v0.26.3 due to scaling limitations in the latest version and remove its alert manager component to disable rps-based scaling. The system setup parameters are listed in Table \ref{table:system parameters}.

\begin{table}[t!]
	\caption{Parameters for System Setup}
	\label{table:system parameters}
	\centering
	\resizebox{0.8\linewidth}{!}
	{\begin{tabular}{| l | l|}
        \hline
        \textbf{Parameter Name} & \textit{\textbf{Value}}     \\ \hline
            MicroK8s version      & v1.27.2                     \\ \hline
            OpenFaaS Gateway version        & v0.26.3                      \\ \hline
            Nodes                   & 5                          \\ \hline
            OS                      & Ubuntu 18.04 LTS            \\ \hline
            vCPU                    & 4,8,12,16                           \\ \hline
            RAM                     & 16,32,64,48 GB                       \\ \hline
            Workload                & Matrix Multiplication ({$m \times m$}) \\ \hline
            m                       & 10(small), 100(medium), 1000(large)     \\ \hline
            CPU, memory, timeout     & 150 millicore, 256 MB, 10 seconds  \\ \hline
	\end{tabular}}
    \vspace{-3.5mm}
\end{table}

\begin{figure}[!t]
	\centering 
	\includegraphics[width=0.9\linewidth]{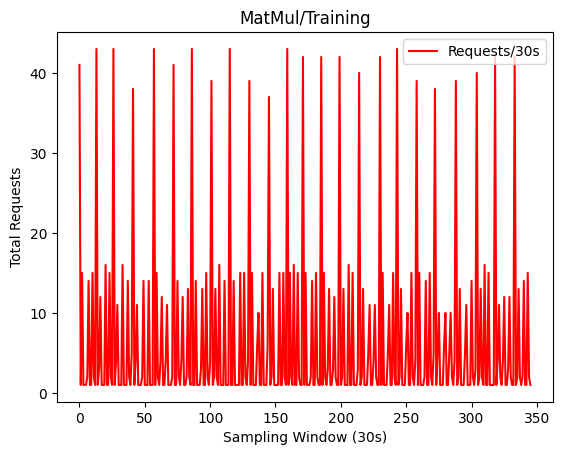}
	\vspace*{-3.5mm}
	\caption{Workload for Matrix Multiplication function}
	\label{fig:workload}
	\vspace*{-4.5mm}
\end{figure}

As FaaS is beneficial for short-running, single-purpose functions that require few resources, we consider \textit{matrix multiplication} function with three different input sizes $small, medium, large$-$(10, 100, 1000)$ and configure it with 150/256 millicore/MB resources approximately as AWS Lambda offering and a maximum timeout of 10 seconds. Additionally, we generate the user workload using the Hey \cite{hey} load generator tool, a lightweight load generator written in Go language. For the workload we leverage an open-sourced, 14-day function trace \cite{shahrad2020serverless} by Azure functions, Fig. \ref{fig:workload}, that largely represents an invocation behaviour of a production-ready application function running on a serverless platform. Although it appears stationary due to its repetitive nature, it is representative of real cloud invocation patterns with relevant variations for scaling decisions. Since the Poisson distribution has been shown to approximately sample online user behaviour, request inter-arrival times are sampled from it. Prometheus service is configured with relevant discovery and target points to regularly scrape metrics from OpenFaaS gateway, function instances and Kubernetes API server.

As discussed in Sec. \ref{section4}, the agent assesses the function demand during a sampling window of 30 seconds for a single episode of 5 minutes. Based on the deployed infrastructure capacity, we fix the maximum function instances as 24 in isolation, to reduce the performance interference. Since frequent scaling can result in resource thrashing, we explore scaling actions within a range of 2 instances, i.e., $a_t \in \{-2, -1, 0, 1, 2\}$, avoiding resource wastage during acquisition and release of function instances. Further, the observation space is composed of the throughput ratio $\phi_t$ $\in$ $[0, 100]\%$, number of function instances $n_t \in [1, 24]$ and resource utilisation (CPU, $c_t$ and memory, $m_t$) $\in [0, 2]*100\%$ that contributes towards over-burdened CPU and out-of-memory scenarios. The LSTM-based PPO agent takes advantage of a single LSTM layer of 256 units and is integrated with both Actor and Critic networks with identical network architectures having 2 fully connected MLP layers of 64 neurons each, i.e., in[64,64] and out[64,64].

\begin{table}[t!]
	\caption{RL Environment and Network Parameters}
	\label{table:rl parameters}
	\centering
	\resizebox{0.8\linewidth}{!}
	{\begin{tabular}{| l | l|}
        \hline
        \textbf{Parameter} & \textit{\textbf{Value}} \\ \hline
        $N$                & 24                       \\ \hline
        $t$                & 30 seconds                     \\ \hline
        $\tau_t$              & (0 - 10) seconds          \\ \hline
        $\phi_t$            &   (0 - 100) \%        \\ \hline
        $q_t$               &   \{0,\dots,Q\} requests    \\ \hline
        $n_t$               &   \{1,\dots,$N$\} functions       \\ \hline
        $c_t$                 &   (0 - 2) *100\%          \\  \hline
        $m_t$               &    (0 - 2) *100\%          \\  \hline
        $a_t$               &   \{-2, -1, 0, +1, +2 \}    \\  \hline
        LSTM Network       & Layer 1(256 cells)         \\  \hline
        Actor Network     & Layer 1(64 cells), Layer 2(64 cells)  \\ \hline
        Critic Network     & Layer 1(64 cells), Layer 2(64 cells)  \\ \hline
	\end{tabular}}
    \vspace{-3.5mm}
\end{table}

\subsection{Experiments}

\begin{figure*}[!htbp]
    \centering
    \begin{minipage}[t]{0.33\textwidth}
        \centering
        \vspace{0pt}
        \includegraphics[width=0.9\textwidth]{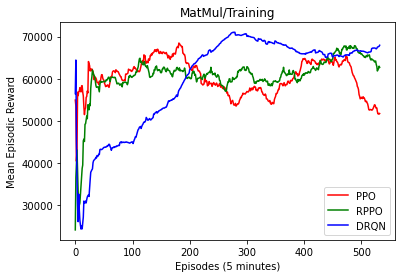}
        \label{fig:train_rewards}
    \end{minipage}
    \begin{minipage}[t]{0.33\textwidth}
        \centering
        \vspace{0pt}
        \includegraphics[width=0.9\textwidth]{ 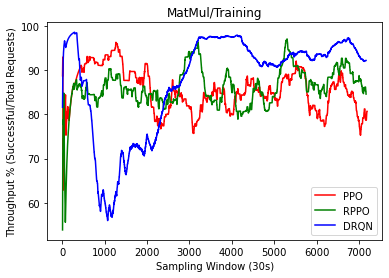}
        \label{fig:throughput_training}
    \end{minipage}
    % \footnotesize{(a)\hspace{250pt}(b)}\\
    \begin{minipage}[t]{0.33\textwidth}
        \centering
        \vspace{0pt}
        \includegraphics[width=0.9\textwidth]{ 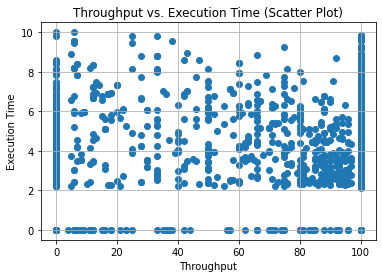}
        \label{fig:ppo_exec_th}
    \end{minipage}
    \footnotesize{(a)\hspace{150pt}(b) \hspace{150pt} (c)}\\
    \begin{minipage}[t]{0.33\textwidth}
        \centering
        \vspace{0pt}
        \includegraphics[width=0.9\textwidth]{ 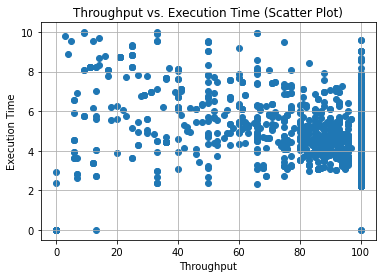}
          \label{fig:rppo_exec_th}
    \end{minipage}
    % \footnotesize{ (c)\hspace{250pt}(d) }
    \begin{minipage}[t]{0.33\textwidth}
        \centering
        \vspace{0pt}
        \includegraphics[width=0.9\textwidth]{ 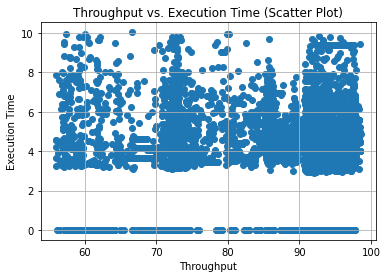}
          \label{fig:drqn_exec_th}
    \end{minipage}\\
    \footnotesize{ (d) \hspace{150pt} (e) }\\
    \caption{(a) Mean episodic reward, (b) Throughput, (c) Throughput vs Execution time - PPO, (d) Throughput vs Execution time - RPPO and (e) Throughput vs Execution time - DRQN}
    \label{Fig:Training}
     \vspace{-4.5mm}
\end{figure*}

Function autoscaling is a continuous and non-episodic process, however, we set an episode based on the default scaling window of 5 minutes by Kubernetes' horizontal scaling mechanism. To demonstrate the effectiveness of recurrency in autoscaling tasks, we chose a workload with varying resource requirements at different sampling windows. After careful consideration of network parameters and sensitivity analysis, listed in Table \ref{table:rl parameters}, the DRL agent is trained for more than 500 episodes to determine a scaling policy to maximise the throughput while using minimal resources. The agent is expected to retain workload information and perform in accordance with the received feedback. Further, the agent is evaluated against a state-of-the-art, PPO-based autoscaling agent, with the same the Actor/Critic network architecture, (Table \ref{table:rl parameters}) as the RPPO agent, i.e., having 2 MLP layers with 64 neurons each. In addition to it, we evaluate a DRQN agent that integrates a LSTM layer (256 cells) with regular off-policy Deep Q-Network (DQN), and has 2 MLP layers with 128 neurons, each for target network and q-network. Fig. \ref{Fig:Training} shows the training results of these competing approaches in terms of mean episodic rewards. The rewards are given as per Eq. \ref{eq:reward}, and it is evident that the mean episodic reward for PPO (60190) begins to diminish after ~400 episodes as compared to LSTM-PPO(RPPO) (60540) agent. Additionally, a similar pattern is visible for the throughput of RPPO and PPO approaches, Fig. \ref{Fig:Training}(b) where PPO struggles to keep a higher success rate by provisioning more functions. Also, we observe that mean episodic reward for the DRQN (59564) approaches that of RPPO while exploring the search space and gradually serves more workload successfully, Fig. \ref{Fig:Training}(e), but closely tailing the trend of other approaches, Fig. \ref{Fig:Training}(b). As mentioned in section \ref{section5}, matrix multiplication is performed for three input sizes - $small, medium, large$ and similar input randomness is followed for competing approaches that are evident in execution time (~$3.7$ and $4$ seconds) of successful requests in Fig. \ref{Fig:Training}(c), (d) and (e).

\begin{figure*}[!htbp]
    \centering
    \begin{minipage}[t]{0.33\textwidth}
        \centering
        \vspace{0pt}
        \includegraphics[width=0.9\textwidth]{ 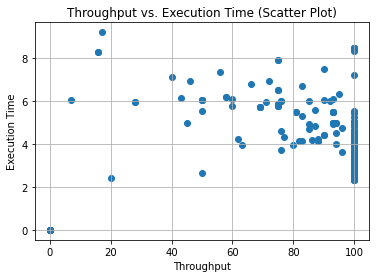}
        \label{fig:eval-ppo-throughput-exec}
    \end{minipage}
    \begin{minipage}[t]{0.33\textwidth}
        \centering
        \vspace{0pt}
        \includegraphics[width=0.9\textwidth]{ 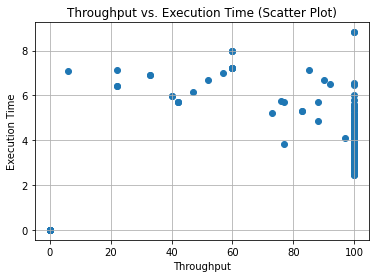}
        \label{fig:eval-rppo-throughput-exec}
    \end{minipage}
    % \footnotesize{(a)\hspace{250pt}(b)}\\
        \begin{minipage}[t]{0.33\textwidth}
        \centering
        \vspace{0pt}
        \includegraphics[width=0.9\textwidth]{ 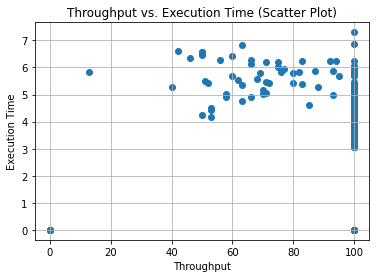}
        \label{fig:eval-drqn-throughput-exec}
    \end{minipage}\\
    \footnotesize{(a)\hspace{170pt}(b) \hspace{150pt}(c)}\\
    \begin{minipage}[t]{0.33\textwidth}
        \centering
        \vspace{0pt}
        \includegraphics[width=0.9\textwidth]{ 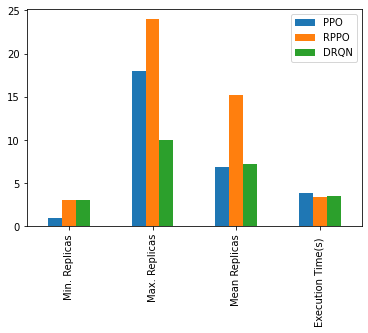}
        \label{fig:eval-replica-count}
    \end{minipage}
    \begin{minipage}[t]{0.33\textwidth}
        \centering
        \vspace{0pt}
        \includegraphics[width=0.9\textwidth]{ 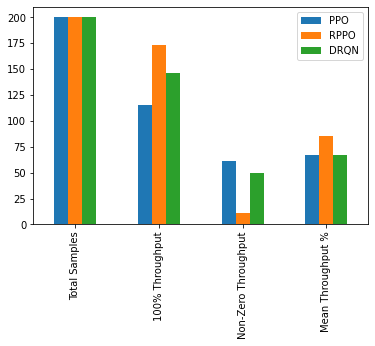}
          \label{fig:eval_sample_count}
    \end{minipage}\\
    \footnotesize{ (d)\hspace{150pt}(e)}\\
    \caption{(a) Throughput vs Execution time - PPO, (b) Throughput vs Execution time - RPPO, (c) Throughput vs Execution time - DRQN, (d) Function Replicas and Execution time, and (e) Throughput Comparison }
    \label{Fig:EvaluationPR}
     \vspace{-4.5mm}
\end{figure*}

We evaluate the agents for 200 sampling windows and present the results in Fig. \ref{Fig:EvaluationPR}. Out of the 200 sampling windows, RPPO based autoscaling agent performed 18\% better in terms of throughput, while having an average of 85\% mean success ratio as compared to 67\% of the PPO agent. On the other hand, the DRQN agent fell short to serve 22\% of the workload with a mean success rate of 66\% as compared to RPPO agent. In serving the evaluated workload, the RPPO agent utilised at least 8.4\% more resources than the PPO agent and improved average execution time (seconds) by 13\%, while it utilised at least 8\% more resources than DRQN and slightly improved the average execution time (seconds) by 2.6\%. Although the DRQN agent tries to capture sequential dependency of the workload, we suspect it fails to explore the search space and only exploits minimal replica count. Hence, as evident in Fig. \ref{Fig:EvaluationPR}(d), the DRQN agent keeps utilising lesser function resources. This agent behaviour is in-line with training results where it could serve better with less requests, thus receiving higher reward for that sampling window and eventually, accumulating higher episodic reward and throughput percentage.  

We also assess the effectiveness of our approach against a default commercial scaling policy, CPU threshold-based horizontal scaling. Kubernetes-based serverless platforms like OpenFaaS \cite{openfaas} and Kubeless \cite{kubeless} can leverage underlying resource-based scaling, known as horizontal pod autoscaling (HPA) implemented as a control loop that checks for target metrics to adjust the function replicas. HPA has a pre-configured query period of 15 seconds to control deployment based on target metrics like average CPU utilization. Therefore, the HPA controller fetches the specific metrics from the underlying API and empirically calculates the number of desired functions. However, the controller is unaware of workload demand and only scales after a 15-second metric collection window. The expected threshold for function average CPU utilisation is set to be 75\% with maximum scaling up to 24 instances. Therefore, whenever the average CPU utilisation of a function exceeds the fixed threshold, new function instances are provisioned. Also, HPA has a 5-minute down-scaling window during which resources are bound to function irrespective of incoming demand, representing potential resource wastage. 
Similarly, we compare our scaling methods with another metric-based autoscaling supported by OpenFaaS based on request-per-second processed. It is also implemented as a control loop and watches for processed requests per second (rps) and raises an alert if rps is above 5 for 10 seconds (default). Therefore, it is worthwhile to analyse the performance of the DRL-based agent against HPA that reserves enough resources for either idle time or low resource utilisation. 

\begin{figure*}[!htbp]
    \centering
    \begin{minipage}[t]{0.33\textwidth}
        \centering
        \vspace{0pt}
        \includegraphics[width=\textwidth]{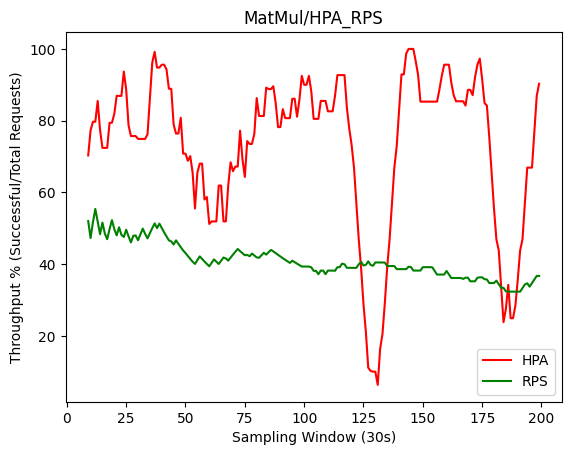}
        \label{fig:hpa_thr}
    \end{minipage}
    \begin{minipage}[t]{0.33\textwidth}
        \centering
        \vspace{0pt}
        \includegraphics[width=\textwidth]{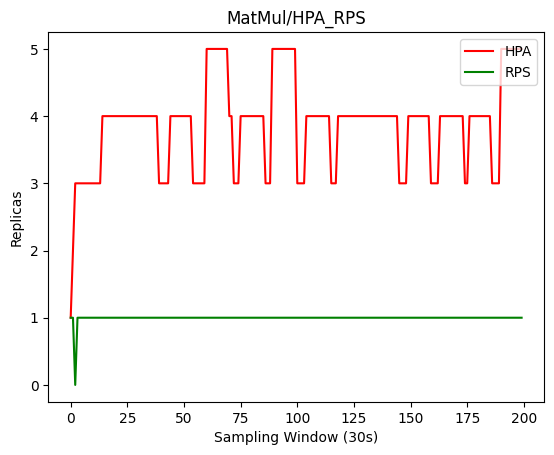}
        \label{fig:hpa_repl}
    \end{minipage}\\
    \footnotesize{(a)\hspace{150pt}(b)}\\
    \caption{(a) Throughput - HPA vs RPS, (b) Replica use - HPA vs RPS}
    \label{Fig:hparps}
     \vspace{-4.5mm}
\end{figure*}

The results for both threshold-based scaling are presented in Fig. \ref{Fig:hparps}, and both approaches struggle to keep up with the incoming workload. The rps could only manage to serve 50\% of incoming load at any sampling window while only using a single instance. This happens as a single request takes approximately 4 seconds to process, and rps never goes beyond the set threshold, failing the majority of requests. On the other hand, HPA could serve 80\% of incoming load on average, but fluctuates due to its set cooldown period. Although HPA tries to scale its resources to 5 replicas, its performance is degraded by 35\% against RPPO and similarly, rps degrades throughput performance by 58\%.

\subsection{Discussion}
\label{sub:discussion}

Autoscaling is an essential feature of cloud computing and has been identified as a potential research gap in serverless computing models. As compared to service-oriented architectures where the services are always running, FaaS functions run for shorter duration and release resources, if unwanted. Hence, an adaptive scaling solution is critical in handling complex workloads for these small and ephemeral functions. Thus, we investigate a DRL-based autoscaling agent, LSTM-PPO, to work in complex FaaS settings and utilise relevant environmental information to learn optimal scaling policies. We train and evaluate the proposed solution against a state-of-the-art on-policy PPO approach, alongside commercial default, and infer that LSTM-PPO is able to capture environment dynamics better and shows promising results. Although, we argue that real-time systems are hard to model and transparent to a certain degree and that RL approaches can analyse these uncertainties better. There are certain points to remember associated with the appropriateness and application of RL methods to real systems. 

We model function autoscaling in FaaS as a model-free POMDP and leverage monitoring tools, like Prometheus, to collect the function-related metrics and apply model-free RL methods to learn the scaling policy. In general, RL algorithms are expensive in terms of {data, resource and time}, where an agent interacts with the modelled environment and acquire relevant information over multiple episodes that signify a higher degree of exploration. Although, as showcased through results, the proposed RL approach took more than 500 episodes ($>$6000 sampling windows) to slightly improve the performance over baselines, RL methods in real-time systems are considerably expensive following stringent optimization goals.

The current training time has an episode of 5 minutes that consists of 10 iteration windows or epochs of 30 seconds each, where a decision to scale is taken by the agent and feedback is calculated for learning. This duration of an episode is chosen keeping in mind the minimum resource scaling and cooldown time of Kubernetes-based serverless platforms and an industry insight \cite{straesser2022not} for taking scaling actions in production environments. In addition to these settings, an agent training could further be affected by the invocation pattern and set of actions to be explored. In the current work, the agent explores 5 discrete actions that follow a conservative approach to avoid resource thrashing while scaling function instances. In a particular state, an agent could take all the possible actions from the action space and would be penalised for an infeasible action. This static behaviour of action modelling elongates the training process since the agent explores infeasible actions in a state and only learns from negative experience. To overcome this, an action masking technique could be integrated that prevents the agent to take certain infeasible actions in particular states, based on defined rules like the total number of function instances to remain within function limits. Therefore, different functions do not necessarily show similar behaviour for training and realised quality of results under similar settings. 

The proposed DRL method is a composition of two different neural network techniques, recurrent and fully-connected layers, and these models are known to be sensitive to respective hyper-parameters or application/workload context. Therefore, configuring hyper-parameters can also be an intensive task in real-world settings. Additionally, the proposed agent analyzes individual workload demand for a particular function, the learning cannot be generalized to other functions with different resource requirement profiles and therefore requires individual training models to be commissioned. However, techniques like transfer learning or categorising functions with similar resource and workload profile to use a trained agent as a starting point could be explored. Moreover, these agents could be deployed in similar fashion to tools like AWS Compute Optimiser \cite{AWSComputeOpt} to gradually obtain experiences and build models with high confidence, from real-time data before making any recommendation/autonomous decision.

Furthermore, the agent is trained for a limited number of episodes, approximately 500 episodes and evaluated, but the chances of exploring are limited. Therefore, the agent expects to be guided by its actor-critic network policies in making informed decisions. Additionally, the agent utilizes resource-based metrics that affect the cold starts, so the availability of relevant tools and techniques to collect instantaneous metrics is essential \cite{straesser2022not} in reducing the observation uncertainty. Also, the respective platform implementation, such as metric collection frequency, function concurrency policy, and request queuing, can extend support to the analyses.
Hence, based on performance evaluation results and discussion, we can adequately conclude that the proposed LSTM-PPO agent successfully performs at par with competing policies for given workload and experimental settings. 

\section{Conclusions and Future Work}
\label{section6}
The FaaS model executes the piece of code inside a container, known as a function and prepares new function containers on demand. FaaS platforms usually support threshold-based autoscaling mechanisms like CPU utilisation to cope with incoming demand and heuristically create more functions. These methods do not consider any system complexity or workload characteristics for scaling and therefore result in sub-optimal scaling policies. Therefore, an adaptive autoscaling mechanism is required to analyse the workload and system uncertainty to optimally scale resources while improving system throughput.

In this work, we investigated a recurrent RL approach for function autoscaling and presented results against a state-of-the-art PPO algorithm and commercially applied threshold-based autoscaling. We perform our analyses for matrix multiplication function and utilise an open-source function trace by Azure \cite{shahrad2020serverless}. The experimental multi-node cluster was set up on the MicroK8s distribution and took advantage of the OpenFaaS serverless framework. We presented evidence of modelling real-time FaaS environments as partially observable and application of recurrent networks to model-free RL algorithms to maximise the objective. We evaluate our proposed technique after training of more than 500 episodes and successfully validate our hypothesis that recurrent techniques capture the system dynamicity and uncertainty to give better autoscaling policies. In our evaluation setting, experiments show that RPPO improved system throughput by 18\%, 22\%, 35\% and 58\% in comparison to PPO, DRQN, HPA and rps scaling policy, respectively.  

As part of future work, we will extend our analysis to different functions and workload types to examine the effect of POMDP modelling. We further plan to experiment with other on-policy and off-policy RL methods like TD3, to expedite the learning process due to their sample efficiency. The proposed methods are dependent on the metric collection process for observing system states which can act as bottlenecks and single points of failure \cite{straesser2022not}. Therefore, we plan to investigate distributed metric collection and agent learning to avoid single-point-failure and improve learning and sample efficiency for estimating optimal function autoscaling policies.

\textbf{Software Availability:} Our environment setup code and algorithms we implemented for OpenFaaS can be accessed from: \url{https://github.com/Cloudslab/DRe-SCale} 
\section*{Acknowledgments}
This research is partially supported by ARC Discovery Project and Melbourne Scholarships. We thank the Melbourne Research Cloud for providing the infrastructure used in the experiments. 

\bibliography{reference}
\bibliographystyle{ieeetr}

\end{document}